\documentclass[11pt,a4paper]{article}
\usepackage[margin=2.5cm]{geometry}
\usepackage{graphicx}
\usepackage{booktabs}
\usepackage{multirow}
\usepackage{amsmath}
\usepackage{amssymb}
\usepackage{hyperref}
\usepackage{xcolor}
\usepackage{float}  % 用于强制图片位置

\title{Design and Empirical Study of a Large Language Model-Based Multi-Agent Investment System for Chinese Public REITs}
\author{Zheng Li\\[0.3em] \small lizheng202601@163.com}
\date{}

\begin{document}

\maketitle

\begin{center}
\section*{Abstract}
\begin{minipage}{0.95\textwidth}
This study addresses the low-volatility Chinese Public Real Estate Investment Trusts (REITs) market, proposing a large language model (LLMs) driven trading framework based on multi-agent collaboration. The system constructs four types of analytical agents—announcement, event, price momentum, and market—each conducting analysis from different dimensions; then the prediction agent integrates these multi-source signals to output directional probability distributions across multiple time horizons; then the decision agent generates discrete position adjustment signals based on the prediction results and risk control constraints, thereby forming a closed loop of "analysis–prediction–decision–execution." This study further compares two prediction model pathways: for the prediction agent, directly calling the general-purpose large model DeepSeek-R1 versus using a specialized small model Qwen3-8B fine-tuned via supervised fine-tuning and reinforcement learning alignment. In the backtest from October 2024 to October 2025, both agent-based strategies significantly outperformed the buy-and-hold benchmark in terms of cumulative return, Sharpe ratio, and maximum drawdown. The results indicate that the multi-agent framework can effectively enhance the risk-adjusted return of REITs trading, and the fine-tuned small model performs close to or even better than the general-purpose large model in some scenarios. The project code and data are open-sourced; details can be found at: \url{https://github.com/adennng/REITs-MAS-LLM}
\end{minipage}
\end{center}

\section{Introduction}
In recent years, the application of large language models (LLMs) in the financial field has gradually expanded from "text understanding" to the complete chain of "reasoning–decision–execution." One line of research focuses on the construction and alignment of financial domain-specific LLMs, improving model reliability in financial Q\&A, information extraction, and decision reasoning through financial corpus pre-training, instruction fine-tuning, and preference alignment. For example, works such as FinGPT, BloombergGPT, and PIXIU are frequently used as representative base models and evaluation benchmarks for financial LLMs \cite{yang2023fingpt,wu2023bloomberggpt,xie2023pixiu}. At the same time, an increasing number of studies emphasize that trading tasks require not only language understanding but also the formation of auditable, executable decision logic amidst multi-source information, market noise, and uncertainty. Therefore, "multi-agent collaborative structures" and "post-training optimization (fine-tuning/RL)" have gained significant attention \cite{xiao2025trading,tian2025tradinggroup}.

\subsection{Structured Trading Systems with Multi-Agents}
In real trading processes, information sources are naturally diverse (news, announcements, macroeconomics, technical indicators, etc.). A single model often struggles to cover the entire chain of "retrieval–analysis–attribution–risk control–execution" simultaneously. Therefore, multi-agent systems (MAS) have become an important form for the systematic implementation of financial LLMs. Related research typically simulates institutional decision-making processes through role division (e.g., analyst/trader/risk controller/memory module) and structured communication (debate, summarization, reflection), thereby improving reasoning quality and reducing hallucination risks. For example, TradingGroup proposed a trading MAS composed of multiple collaborative agents and a risk management module, introduced a self-reflection mechanism for multi-level self-correction, and generated instruction data suitable for fine-tuning through a data synthesis and automatic annotation pipeline \cite{tian2025tradinggroup}.

Parallel to the "multi-role collaboration" approach, another line of thought explicitly decomposes the trading system into subtasks such as "direction judgment" and "position/size management" to enhance the controllability of risk control. FinPos explicitly points out that position management in real markets brings higher risk exposure. It proposes a multi-agent decision-making system to decouple directional decisions (buy/sell/hold) from sizing decisions (trade volume/position adjustment) and achieves controllable risk through upper-limit constraints and risk-aware prompts \cite{liu2025finpos}.

\subsection{Fine-tuning and Reinforcement Learning Post-Training}
Besides system structure, how to enable LLMs to produce stable and usable strategies in sequential decision-making tasks like trading is another key research direction. Numerous works point out that directly using general-purpose LLMs often only generates "seemingly reasonable" textual analysis but struggles to consistently produce robust, executable decisions under risk constraints in market backtests. Therefore, it is necessary to combine supervised fine-tuning (SFT) and reinforcement learning post-training (RL/RFT) to align with trading objectives. Trading-R1 regards "structured reasoning required for trading" as a core capability. It constructs a large-scale financial reasoning dataset and adopts a "first SFT, then phased RL training from easy to difficult" approach to make model outputs more structured and evidence-based, aligning decisions with trading principles such as volatility adjustment \cite{xiao2025trading}. It also systematically discusses RLHF/preference optimization family methods (e.g., PPO, DPO, GRPO, etc.), emphasizing the importance of multi-stage RL training and reward design for stable alignment \cite{xiao2025trading,deepseek2025}.

Another line of work more directly treats LLMs as policy networks and introduces environmental returns into the update process via policy gradient methods. FLAG-TRADER proposes integrating LLM language reasoning with gradient-driven RL optimization, emphasizing that parameter-efficient fine-tuning and post-training methods like PPO can map natural language decisions into optimizable trading strategies, thereby improving adaptability across market environments \cite{yu2025flag}. Notably, some studies combine "multi-agent structure" with "training data construction": The data synthesis and automatic annotation pipeline in TradingGroup aims to continuously produce high-quality trading process samples usable for fine-tuning, narrowing the generalization gap caused by financial LLMs' lack of real trading experience \cite{tian2025tradinggroup}.

\subsection{Positioning and Contributions of This Study}
Building upon the above research, this paper, targeting the trading scenario of Chinese Public Real Estate Investment Trusts (REITs) (a market with far lower volatility than stocks), proposes a large model-driven closed-loop framework of "multi-dimensional interpretation $\rightarrow$ multi-horizon directional probability prediction $\rightarrow$ discrete position adjustment signals":

1) At the information layer, driving factors are decomposed into four types of agents: announcement, event, price momentum, and market macro.

2) At the prediction layer, the above multi-source factors are fused into multi-horizon directional probability judgments to capture holding period differences.

3) At the execution layer, prediction results are further mapped into discrete position adjustment intensity signals. The system completes target position calculation and trade execution, forming a closed loop of "analysis–prediction–decision–execution."

Furthermore, for the prediction layer, this study provides two implementation pathways, employing different models:

(i) General-purpose reasoning large model pathway: directly calling a large-scale general-purpose model like DeepSeek-R1.

(ii) Specialized small-scale LLM pathway: training on historical data (12 months from October 2023 to October 2024) to perform SFT and RL on a small-scale LLM (e.g., Qwen3-8B) to enhance its analytical capability in specific scenarios. This is to observe whether the fine-tuned small-scale LLM's performance can match or even exceed the predictive and analytical ability of the large-scale general-purpose LLM (DeepSeek-R1).

For experimental evaluation, this study uses data from October 2024 to October 2025 (12 months) for backtesting. The results show that both strategies based on the multi-agent framework significantly outperform the traditional "Buy \& Hold" strategy in terms of cumulative return rate (CR), Sharpe ratio (Sharpe), and maximum drawdown (MDD).

\section{System Architecture}
In this study's public REITs multi-agent investment system, the announcement, event, price momentum, and macro market agents generate structured analysis results from four perspectives: information disclosure, operational events, technical behavior, and macro environment, respectively. The prediction agent and decision agent are located in the system's "fusion–execution" layer: the former is responsible for unifying multi-source information into quantifiable directional predictions, and the latter maps the prediction results into position adjustment actions for single-asset accounts, forming a closed loop of "analysis–prediction–decision–execution."

\subsection{Price Momentum Agent}
The Price Momentum Agent characterizes the market behavior of public REITs from the perspectives of price trends, technical patterns, and volume structure. Its goal is to transform scattered technical signals in a low-volatility market into structured conclusions usable for multi-agent fusion. This module relies solely on observable price and volume data and is characterized by reproducible indicator calculations and constrained reasoning outputs.

\subsubsection{Technical Indicator Calculation and Technical State Construction}
The system calculates and aggregates multiple types of technical indicators on price and volume to form a unified characterization of "trend–momentum–volatility–volume–structural levels":

a) Trend structure: MA5/10/20/60 and their deviations, with discretized judgments on moving average alignments (bullish alignment/bearish alignment/chaotic). It also calculates 1/5/20/60-day price change rates to capture multi-period trend consistency.

b) Momentum signals: RSI(6/12/24) with state labeling (overbought/oversold/normal), MACD (DIF/DEA/Histogram) identifying key turning signals like golden cross/death cross, and calculation of 10-day momentum values to capture momentum strength changes.

c) Volatility and range structure: Bollinger Bands (20,2$\sigma$) position (breaking upper band/breaking lower band/middle/biased upward/biased downward) and 20-day volatility level, with simplified ATR as a short-term volatility reference.

d) Volume and price-volume relationship: Calculation of volume MA5/10/20, volume ratio (daily volume / 5-day average volume), and output of price-volume structure labels based on rules like "price up volume up/price up volume down/price down volume up/price down volume down" to distinguish trend confirmation and momentum decay.

e) Key structural levels (support/resistance): Based on rolling window high/low points (5/10/20/60) and candidate sets like MA and Bollinger Bands, automatically selecting the most representative support and resistance levels to provide interpretable price references for subsequent "trend continuation/false breakout/pullback risk."

Additionally, the system counts consecutive up/down days, number of up/down days in the recent 20 days, average/maximum amplitude in the recent 20 days, and specifically outputs the daily price change rates for the last 5 trading days.

\subsubsection{Dynamic Volatility Threshold and "Sideways" Modeling}
To adapt to the low-volatility market characteristics of Chinese Public Real Estate Investment Trusts (REITs) market, a daily dynamic volatility threshold is introduced to judge whether price movements are statistically significant. This threshold is defined as:

\[
\theta_t = \max \Big( q_L,\; \min \big( \sigma_t \cdot m_t,\; q_U \big) \Big)
\]

where:

* $\sigma_t$ is the standard deviation of returns over the most recent $N_v$ trading days up to trading day $t$ (short-term historical volatility), used to characterize the instantaneous volatility level of the current market.

* $m_t$ is the \textbf{adaptive multiplier}, adjusted based on the ratio of short-term to long-term volatility:

\[
m_t =
\begin{cases}
m_0 \cdot a_{\text{high}}, & \text{if } \frac{\sigma_t^{(\text{short})}}{\sigma_t^{(\text{long})}} > \tau_{\text{high}} \\
m_0 \cdot a_{\text{low}}, & \text{if } \frac{\sigma_t^{(\text{short})}}{\sigma_t^{(\text{long})}} < \tau_{\text{low}} \\
m_0, & \text{otherwise}
\end{cases}
\]

where $m_0$ is the base multiplier, $\sigma_t^{(\text{short})}$ and $\sigma_t^{(\text{long})}$ represent short-term and long-term historical volatility, respectively, $\tau_{\text{high}}$ and $\tau_{\text{low}}$ are thresholds for high and low volatility states, and $a_{\text{high}}$ and $a_{\text{low}}$ are corresponding adjustment coefficients.

* $q_L$ and $q_U$ are the lower and upper quantiles of the absolute return distribution over the past $N_b$ trading days, used to prevent the threshold from being pulled too small or too large by extreme volatility during abnormal periods.

In actual parameter settings, the short-term volatility window $N_v$ is about 30 days, the quantile boundaries $(q_L, q_U)$ typically take the 30\% and 70\% quantiles, the adaptive volatility judgment thresholds $(\tau_{\text{high}}, \tau_{\text{low}})$ are approximately 1.4 and 0.7, respectively, and the base multiplier $m_0$ and adjustment coefficients $(a_{\text{high}}, a_{\text{low}})$ are used to characterize reasonable amplification or contraction ratios in the low-volatility environment of public REITs.

The dynamic threshold $\theta_t$ obtained through this mechanism defines the boundary for "statistically significant price movement" at the single-day level:

\[
|r_t| > \theta_t \Rightarrow \text{trending movement}, \qquad
|r_t| \le \theta_t \Rightarrow \text{sideways}
\]

where $r_t$ is the daily return.

To maintain consistency in the definition of "sideways" across different prediction horizons, the system extends the single-day threshold $\theta_t$ to multi-day intervals via the square root of time rule based on the volatility scaling relationship of prices approximating a random walk:

\[
\varepsilon_{5} = \sqrt{5}\cdot \theta_t, \qquad
\varepsilon_{20} = \sqrt{20}\cdot \theta_t
\]

where:

* $\varepsilon_{5}$ is the \textbf{sideways threshold for T+5 (next 5 trading days)}.

* $\varepsilon_{20}$ is the \textbf{sideways threshold for T+20 (next 20 trading days)}.

Therefore, the multi-horizon "sideways" determination rules are:

* \textbf{T+1 (single day)}:

\[
|r_{t+1}| \le \theta_t \;\Rightarrow\; \text{sideways}
\]

* \textbf{T+5 (5-day cumulative)}:

\[
\left\|\sum_{i=1}^{5} r_{t+i}\right\| \le \varepsilon_{5} = \sqrt{5}\cdot\theta_t
\;\Rightarrow\; \text{sideways}
\]

* \textbf{T+20 (20-day cumulative)}:

\[
\left\|\sum_{i=1}^{20} r_{t+i}\right\| \le \varepsilon_{20} = \sqrt{20}\cdot\theta_t
\;\Rightarrow\; \text{sideways}
\]

Equivalently, this means: if the absolute value of the average daily return over the next $k$ days does not exceed $\theta_t$, the price movement on that time scale is considered sideways absorption rather than trending.

Testing with historical data shows that under this parameter setting, approximately 33\% of trading days have price movement absolute values not exceeding the corresponding dynamic threshold. This makes "sideways" a statistical normal category matching the volatility structure of the REITs market, maintaining scale consistency across T+1, T+5, and T+20.

\subsubsection{LLM-Driven Technical Signal Synthesis}
After completing all technical indicator calculations, the system organizes information such as "trend structure, momentum state, Bollinger Bands position, price-volume relationship, volatility level, support and resistance ranges, and whether recent days have broken through the dynamic sideways threshold" into structured context and inputs it into a large language model (DeepSeek-R1). Under a clear technical analysis framework and terminology constraints, the model performs joint reasoning on multi-dimensional signals, outputting judgments including the current price stage (e.g., trend continuation, consolidation absorption, or trend exhaustion), explanations of key technical phenomena (e.g., whether a valid breakout has formed, whether momentum divergence appears, whether overbought or oversold conditions exist), and the technical implications for future price paths (e.g., trend likely to continue, pullback risk, or more likely to maintain range-bound oscillation). Simultaneously, the model provides potential key price ranges based on support and resistance levels and annotates the most important risk factors in the current structure (e.g., false breakout, insufficient volume, or volatility contraction).

\subsection{Announcement Agent}
The Announcement Agent is responsible for evaluating announcement-driven price impacts for public REITs under recent information disclosures. Its core goal is to judge: given the current price and volatility state, whether the most recently disclosed set of announcements constitutes trading signals with historical statistical significance.

At the data level, a historical impact analysis module is established for specific types of announcements. When this module is invoked, it extracts historical announcements of the same type for that REIT before the analysis date from the database, including announcement summaries, sentiment annotations, and price performance within specific trading days after the announcement release. These historical records are grouped by "positive / neutral / negative" sentiment, and multiple statistics are calculated accordingly, such as short-term probability of price increase, average increase ratios over different windows, and frequency of significant fluctuations under positive or negative scenarios, thereby forming a historical price reaction profile for that announcement type.

In the real-time phase, the system collects all announcements released within the 7 natural days preceding the analysis date and filters out key announcement types requiring historical comparison. For these announcements, the system automatically calls the corresponding historical impact analysis module to obtain their historical statistical performance.

Simultaneously, the system obtains the price trend of that REIT over the recent 5–20 trading days. This is used to judge whether announcements that are typically influential historically are more likely to be amplified or absorbed by sideways market conditions in the current market environment.

The above three types of information—

(i) summary of announcements from the past 7 natural days,  
(ii) historical statistical reaction of similar announcements,  
(iii) current price and sideways state

are organized into structured context and input into the large language model (DeepSeek-R1). The model performs comprehensive reasoning, outputting a structured interpretation of the current announcement set, including: whether there exist pricing-significant announcements and their key information, their potential directional impact, and whether that impact is more likely to be amplified or attenuated in the current market condition.

\subsection{Event Agent}
The Event Agent is used to identify the event-driven phase of public REITs: i.e., whether recent news catalysts, changes in underlying asset operations, or quarterly report expectation games have occurred that could affect pricing. The agent's input consists of three types of information: first, market news within a fixed window before the analysis date (focusing only on high-impact news, with higher weight given to news closer to the analysis date); second, the fund's most recent quarterly report and several operational data reports issued after the quarterly report. The system retains summaries, sentiment annotations, and "reasoning" for each report, focusing on whether operations are improving/deteriorating and whether they are above normal levels; third, a quarterly report warning module is used to judge whether the current time point is close to the quarterly report release date and accordingly prompts that other investors may engage in pre-report positioning games.

At the output level, the Event Agent organizes the above information into structured context and inputs it into the large language model (DeepSeek-R1). The model ultimately outputs: key points of the filtered news, recent changes in underlying asset operations, whether there are quarterly report warnings and their investment suggestions, and provides comprehensive analysis and risk alerts.

\subsection{Market Agent}
The Market Agent is responsible for making macro-level judgments on the overall allocation environment of the public REITs market from three core dimensions: interest rate environment, equity market risk appetite, and the valuation state of the REITs market itself. This module does not analyze individual events of single funds but provides a top-down macro and market background constraint for the multi-agent system, answering "whether the current REITs market is in a systemic tailwind or headwind period."

\subsubsection{Multi-layer Market State Construction}
The Market Agent's input comes from four types of market data: CSI REITs Total Return Index, 10-year government bond yield, stock indices (Shanghai Composite Index and CSI Dividend Index), and REITs whole market turnover and market capitalization data. The system first calculates numerous quantitative indicators from these raw time series, including:

* \textbf{REITs Market's Own State}: Price historical quantile of CSI REITs Total Return Index, short/medium/long-term price change rates, moving average structure, RSI, MACD, volatility quantile, proportion of up days, etc.

* \textbf{Interest Rate Environment}: Absolute level of the 10-year government bond yield, 1-year quantile, 20-day change magnitude, deviation from moving averages, and interest rate–REITs correlation.

* \textbf{Equity Market Environment}: Price change rates, RSI, trend status of the Shanghai Composite Index and Dividend Index, and the relative strength of CSI REITs Total Return Index versus the Dividend Index (to capture the capital seesaw effect).

* \textbf{Market Sentiment}: Turnover rate, trading volume, and price-volume relationship of the entire REITs market, to characterize capital activity.

These numerical indicators are then mapped into interpretable state labels (e.g., "interest rate relatively high," "momentum relatively weak," "dividend strong," "turnover rate sluggish," etc.), forming a three-layer data representation: raw indicator layer $\rightarrow$ interpretation layer $\rightarrow$ state descriptions usable for decision-making.

\subsubsection{Four-Quadrant Macro Allocation Framework (Core)}
The core structure of the Market Agent is a four-quadrant model with interest rate trend and equity market state as axes. This serves as the unified coordinate system for all macro judgments in the system.

a) Horizontal Axis: Interest Rate Trend (Valuation Anchor for REITs)

The system discretizes the interest rate environment based on the 20-day change magnitude (in basis points) of the 10-year government bond yield:

* Clearly Downward ($< -20$bp)  
* Slowly Downward ($-20$bp to $-5$bp)  
* Sideways Oscillation ($-5$bp to $+5$bp)  
* Slowly Upward ($+5$bp to $+20$bp)  
* Clearly Upward ($> +20$bp)

This dimension directly characterizes whether the discount rate environment for REITs is in a "tailwind" or "headwind."

b) Vertical Axis: Equity Market State (Capital Seesaw)

The equity market state is jointly determined by the 20-day price change rate and RSI of the Shanghai Composite Index, strictly divided into five categories:

* Bull Market (price change $>5\%$ and RSI $>60$)  
* Oscillation with Strength (price change $0\%$–$5\%$, RSI $50$–$60$)  
* Oscillation (price change $-2\%$ to $2\%$, RSI $40$–$60$)  
* Oscillation with Weakness (price change $-5\%$ to $0\%$, RSI $40$–$50$)  
* Bear Market (price change $< -5\%$, RSI $<40$)

This dimension reflects whether capital is more inclined to flow into equity assets or prefers stable income assets like REITs.

c) Economic Implications of the Four Quadrants

Based on the combination of interest rate trend and equity market state, the system divides the market into four quadrants with clear economic implications:

* \textbf{Quadrant I (Interest Rate Downward + Equity Market Upward)}: REITs valuation is supported by interest rates, but strong equity markets may divert capital; overall "cautiously optimistic."

* \textbf{Quadrant II (Interest Rate Upward + Equity Market Upward)}: Discount rate rising and capital flowing to equities, creating a "double squeeze" on REITs; the most unfavorable environment.

* \textbf{Quadrant III (Interest Rate Downward + Equity Market Oscillation or Downward)}: Valuation benefits from falling interest rates, and capital flows back to stable assets; the best allocation window for REITs.

* \textbf{Quadrant IV (Interest Rate Upward + Equity Market Downward)}: Interest rates suppress valuation, but equity market weakness brings some safe-haven demand; REITs are defensive but with limited returns.

When interest rates or the equity market are in a "sideways" or borderline range, they are classified into a \textbf{transition zone} to avoid overly aggressive judgments when macro signals are unclear.

\subsubsection{LLM-Driven Comprehensive Market Interpretation}
After obtaining the interest rate trend, equity market state, REITs valuation quantile, and market sentiment indicators through quantitative calculation, and completing the four-quadrant positioning, the Market Agent inputs these results into the large language model (DeepSeek-R1) in a three-layer form: structured summary, interpretation layer, and raw indicators.

Within the explicit framework of the "four-quadrant" model, it analyzes the consistency and prioritizes contradictions among the multi-dimensional signals, outputting an interpretation of the current environment in the REITs market. This includes: the current quadrant position and its economic implications, the combined directional influence of interest rates and the equity market on REITs, and whether the overall environment leans more towards an allocation window or a risk-suppression phase. This output then serves as the macro-environmental factor within the system.

\subsection{Prediction Agent}
The Prediction Agent (Direction Predictor) receives the structured analysis from the above four agents along with necessary price context information. It then passes this information to the LLM, which outputs directional probability distributions (up/down/side) and corresponding confidence levels across three time horizons (T+1, T+5, T+20), used to express directional judgments and uncertainty levels for different holding periods.

\subsection{Decision Agent}
The Decision Agent obtains the multi-horizon directional prediction results from the Prediction Agent, along with the current account status and risk control constraints, such as position limits, fixed-amount position adjustment step sizes, position-building phase pace restrictions, and risk control rules. The system then passes the above content to the LLM (DeepSeek-R1), which outputs a discrete position adjustment signal (e.g., close position, reduce position by 40\%, reduce position by 20\%, hold unchanged, increase position by 20\%, increase position by 40\%, increase position to the upper limit). The system side maps this to actual position changes for execution and updates the account information.

\begin{figure}[H]
\centering
\includegraphics[width=\textwidth]{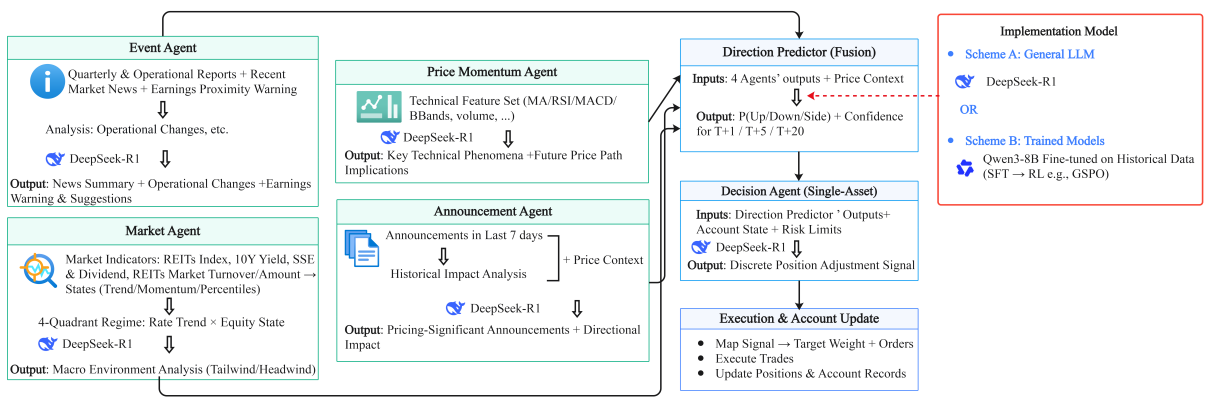}
\caption{Multi-Agent REITs Trading System Architecture Diagram}
\label{fig:architecture}
\end{figure}

\section{Two LLM Choices for the Prediction Agent and the Fine-tuning Process for the Small-Scale LLM}
For the Prediction Agent, this study selects different models for execution: (1) directly calling a general-purpose reasoning large model (DeepSeek-R1); (2) performing SFT and RL on a small-scale LLM (Qwen3-8B) so that prediction and decision are executed by the trained specialized model. This is to observe whether the fine-tuned small-scale LLM's performance can match or even exceed that of the large-scale general-purpose LLM (DeepSeek-R1).

\subsection{Fine-tuning Process for the Small-Scale LLM}
The chosen small-scale large language model is Qwen3-8B. Training adopts a "two-stage paradigm": first, obtaining stable structured output capability through supervised fine-tuning (SFT), followed by reward-based GSPO reinforcement learning to align with real price movements and prediction preferences. The training data source is the real price data of all REITs over the 12 months from October 2023 to September 2024. Considering that the funds selected for the backtest phase are those listed for over one year, the training phase also only selects data for funds that have been listed for more than one year on the respective trading day as training samples.

\subsubsection{Task Definition and Input/Output}
The goal of the Prediction Agent is: for a single REIT, output directional probability distributions (up/down/side) and confidence levels across three holding periods (T+1, T+5, T+20). Its input consists of the structured analysis results provided by the four upstream agents (announcement, event, price momentum, market) along with necessary price context. The output adopts a structured template of "reasoning process + strict JSON."

\subsubsection{True Label and Dynamic Threshold Annotation}
To quantify "up/down/side" in the low-volatility REITs market, as mentioned earlier, the system introduces a dynamic threshold $\theta_t$ and extends it to multi-period thresholds via the square root of time rule $\varepsilon_k=\sqrt{k}\theta_t$ (where $k\in\{1,5,20\}$). The true multi-period return is defined as:

\[
R_k=\frac{P_{t+k}-P_t}{P_t},
\]

and the supervision labels are generated accordingly:

\[
\text{label}_k=
\begin{cases}
\text{up}, & R_k\ge \varepsilon_k\\
\text{down}, & R_k\le -\varepsilon_k\\
\text{side}, & \text{otherwise}
\end{cases}
\]

This annotation mechanism ensures that "sideways" is not a fixed threshold assumption but adapts with market volatility, thereby aligning with the low-volatility market structure of REITs.

\subsubsection{Data Construction: Teacher Model Distillation for Learnable Structured Reasoning}
A "teacher model distillation" approach is used to construct high-quality supervision data: for each trading day, the four-dimensional expert information and the future actual price movement are provided to a teacher model (e.g., DeepSeek-R1), which produces "interpretable reasoning chains + structured output satisfying format constraints." The teacher's output serves as the supervision signal for SFT, while the same input structure is used in the subsequent GSPO stage to generate multiple candidate responses and perform reward optimization.

\subsubsection{Two-Stage Training: SFT Foundation + GSPO Alignment with Real Price Movements}
Stage One (SFT): Autoregressive cross-entropy training is used to first make the model learn to stably output structured conclusions of "reasoning + JSON" and learn the teacher model's reasoning approach:

\[
\mathcal{L}_{\text{SFT}}(\theta)=
-\frac{1}{N}\sum_{n=1}^{N}\sum_{t=1}^{T_n}\log
p_\theta(y_{n,t}\mid x_n, y_{n,<t})
\]

where $y$ covers the complete tokens of the reasoning segment and structured output.

Stage Two (GSPO): Building upon the SFT foundation that already possesses "compliant output capability," reward-based GSPO reinforcement learning is introduced to further align "direction correctness + output compliance." The system computes a normalized reward $R\in[0,1]$ for each model output, with the general form:

\[
R=\alpha\cdot \text{correctness}+\beta\cdot \text{FormatScore},
\qquad \alpha+\beta=1.
\]

where $\alpha,\beta$ are the weights for content score and format score, respectively, emphasizing "directional prediction correctness" as the primary optimization goal while retaining constraints on structured output compliance.

(a) Content Score \textit{correctness}: Whether Directions Across Three Time Horizons Are Correct

The model outputs three directional probabilities $\{p_k(\text{up}),p_k(\text{down}),p_k(\text{side})\}$ for each time horizon $k\in\{1,5,20\}$ and takes the dominant direction:

\[
\hat{y}_k=\arg\max_{c\in\{\text{up,down,side}\}} p_k(c).
\]

Comparing with the true label $y_k$ yields an indicator function:

\[
I_k=\mathbb{I}[\hat{y}_k=y_k],
\]

Then \textit{correctness} is defined as the weighted accuracy:

\[
\text{correctness}=w_1 I_1+w_5 I_5+w_{20} I_{20},
\qquad w_1+w_5+w_{20}=1.
\]

\textit{correctness} is the content score $w_{\text{dir}}$. To adapt to the "within-group normalization" training setup of GSPO/GRPO, the content score only retains directional correctness (avoiding introducing factors like confidence or magnitude that are approximately constant within the same group into the reward, thereby mitigating the side effect of "overconfidence").

(b) Format Score \textit{FormatScore}: Compliance Degree with Structured Output Template and Numerical Constraints

The format score encourages the model to stably output structured results of "`<think>...</think>` + JSON" and constrains key fields to be complete and numerical values reasonable. The composition of \textit{FormatScore} considers: basic format (whether it contains `<think>` tags, whether JSON is parsable, etc.), field completeness (whether JSON contains all necessary fields), numerical constraints (whether probabilities sum to 1 for each horizon, whether dominant probabilities fall within specified intervals, and whether all probabilities are not below minimum thresholds, etc.).

In summary, the GSPO stage, through the joint reward of "direction correctness (main goal) + output compliance (hard constraint)," encourages the prediction model to more directly align with the direction labels corresponding to future real price movements while maintaining structured and usable outputs, thereby achieving a balance between "output usability" and "prediction relevance."

\section{Experiments}

\subsection{Backtest Period and Dataset}
The first batch of Chinese Public Real Estate Investment Trusts (REITs) was listed in June 2021. Considering the relatively small market size in its early development stage, with pricing mechanisms and liquidity expectations not yet mature, overall volatility was relatively pronounced. To minimize the impact of abnormal stage-specific market fluctuations and make the backtest results more representative and robust, this study sets the backtest period from October 2024 to October 2025, totaling 12 months.

Simultaneously, to avoid interference from non-normal factors such as significant price volatility, unstable valuation systems, and drastic liquidity changes often observed in newly issued REITs during their initial listing phase, the sample for this study only includes funds that had been listed for at least one year as of the backtest start date (October 1, 2024). After screening, 28 funds met the criteria. The specific list is detailed in the statistical table later.

The data involved in the experiments are all from public market information, including each fund's daily closing price, trading volume, turnover rate, daily announcements, daily prices of the CSI REITs Total Return Index and stock indices (Shanghai Composite Index and CSI Dividend Index), daily 10-year government bond yield data, important news during the backtest period, etc.

\subsection{Baseline and Evaluation Metrics}
\noindent(1) Backtest Settings

This trading adopts a single-asset independent backtest mode. Each fund corresponds to an independent simulated trading account to individually evaluate the effectiveness of the investment framework on different funds. All accounts start with the same initial capital (RMB 1 million) and are strictly limited to trading only their corresponding fund. When a fund's position is not full, the remaining capital is kept as cash in the account. The transaction cost is set at 0.03\% (0.0003) per trade.

\noindent(2) Strategy Groups and Control Group

Both strategy groups are the execution results of this multi-agent framework. Their core difference lies in the model used by the Prediction Agent:

Strategy A: Prediction task executed by the general-purpose large language model DeepSeek-R1.

Strategy B: Prediction task executed by the fine-tuned small-scale language model Qwen3-8B.

Additionally, a control group strategy is specially set up, namely the classic "Buy \& Hold" strategy. This strategy invests all initial capital into the corresponding fund on the first trading day of the backtest period and then makes no further trades until the end of the backtest period.

\noindent(3) Evaluation Metrics

The selected evaluation metrics are: (1) Cumulative Return Rate (CR): a return metric capturing the total profit generated during the backtest period. (2) Sharpe Ratio (SPR): a risk-adjusted performance metric indicating the excess return per unit of risk. (3) Maximum Drawdown (MDD): a risk metric measuring the most severe peak-to-trough loss.

\section{Results}

\subsection{Overall Performance Comparison}
Table 1 summarizes the average performance of the three strategies across all 28 funds. It can be seen that both strategies based on the multi-agent framework (Agent-DeepSeek-R1 and Agent-Qwen3-8B-FT) significantly outperform the traditional "Buy \& Hold" strategy in terms of cumulative return rate (CR), Sharpe ratio (Sharpe), and maximum drawdown (MDD).

Cumulative Return Rate (CR): The DeepSeek-R1 pathway is slightly higher than the fine-tuned small model pathway (15.50\% vs 13.75\%), and both are clearly higher than the benchmark (10.69\%).

Sharpe Ratio (SPR): The Qwen3-8B-FT pathway is slightly higher than the DeepSeek-R1 pathway (1.77 vs 1.71), and both are far above the benchmark's 0.75. This indicates that the agents do not merely achieve returns by increasing volatility but obtain returns under a more robust risk structure. Notably, the Qwen3-8B-FT pathway, despite having a slightly lower average return, has a slightly higher Sharpe ratio, indicating stronger stability of returns and risk control capability.

Maximum Drawdown Rate (MDD): The average maximum drawdown of both agent strategies is controlled within -5\%, while the average drawdown of the benchmark strategy exceeds -11\%. In extreme cases (minimum value column), the benchmark strategy's maximum drawdown can reach -24.19\%, while the worst drawdowns for the agent strategies are -8.07\% (DeepSeek-R1) and -8.24\% (Qwen3-8B-FT), respectively, showing significantly improved risk resilience.

Both agent pathways jointly exhibit the structural characteristic of "returns not decreasing (even increasing) + drawdowns significantly decreasing."

\subsection{Individual Fund Performance Analysis}
From the perspective of individual funds (detailed in Table 2), both agent strategies outperform the benchmark on the vast majority of funds.

Win rate against Buy\&Hold (CR\%): The Agent-DeepSeek-R1 pathway achieves higher CR on 75\% of the underlying assets; the Agent-Qwen3-8B-FT pathway achieves higher CR on 60.7\% of the underlying assets.

Win rate against Buy\&Hold (Sharpe): Agent-DeepSeek-R1 has superior Sharpe on 89.3\% of the underlying assets; Agent-Qwen3-8B-FT is superior on 85.7\% of the underlying assets.

Win rate against Buy\&Hold (MDD\%): Agent-DeepSeek-R1 has smaller drawdowns on 100\% of the underlying assets; Agent-Qwen3-8B-FT has smaller drawdowns on 96.4\% of the underlying assets.

% Table 1
\subsection*{Table 1: Summary of Overall Strategy Performance Comparison}
\begin{tabular}{llrrr}
\toprule
Metric & Statistic & Agent-DeepSeek-R1 & Agent-Qwen3-8B-FT & Baseline-Buy \& Hold \\
\midrule
\multirow{3}{*}{CR\%} & Mean & 15.50 & 13.75 & 10.69 \\
 & Max & 50.25 & 37.53 & 49.52 \\
 & Min & 0.62 & 0.44 & -13.13 \\
\addlinespace
\multirow{3}{*}{Sharpe} & Mean & 1.71 & 1.77 & 0.75 \\
 & Max & 4.47 & 3.67 & 3.46 \\
 & Min & -0.27 & -0.63 & -1.11 \\
\addlinespace
\multirow{3}{*}{MDD\%} & Mean & -4.09 & -3.46 & -11.12 \\
 & Max & -1.10 & -0.23 & -5.58 \\
 & Min & -8.07 & -8.24 & -24.19 \\
\addlinespace
\bottomrule
\end{tabular}

% Table 2
\subsection*{Table 2: Details of Strategy Performance for Each Fund}
\begin{tabular}{lrrrrrrrrr}
\toprule
 & \multicolumn{3}{c}{Agent-DeepSeek-R1} & \multicolumn{3}{c}{Agent-Qwen3-8B-FT} & \multicolumn{3}{c}{Baseline-Buy \& Hold} \\
Fund Code & CR\% & Sharpe & MDD\% & CR\% & Sharpe & MDD\% & CR\% & Sharpe & MDD\% \\
\midrule
180101.SZ & 17.81 & 2.07 & \textcolor{red}{-3.62} & \textcolor{red}{21.40} & \textcolor{red}{2.65} & -3.96 & 1.61 & 0.09 & -13.04 \\
180102.SZ & 18.23 & 2.55 & -3.01 & \textcolor{red}{24.32} & \textcolor{red}{3.67} & \textcolor{red}{-1.76} & -4.18 & -0.29 & -21.15 \\
180103.SZ & \textcolor{red}{9.57} & 1.17 & -4.15 & 6.77 & \textcolor{red}{1.57} & \textcolor{red}{-1.69} & -1.37 & -0.12 & -11.36 \\
180201.SZ & \textcolor{red}{11.49} & \textcolor{red}{2.01} & -2.28 & 2.95 & 0.58 & \textcolor{red}{-1.49} & 4.72 & 0.46 & -7.16 \\
180202.SZ & \textcolor{red}{5.27} & \textcolor{red}{0.92} & -3.01 & 0.52 & -0.63 & \textcolor{red}{-0.60} & -5.94 & -1.11 & -7.84 \\
180301.SZ & \textcolor{red}{18.28} & \textcolor{red}{2.26} & -3.17 & 10.18 & 2.05 & \textcolor{red}{-3.13} & 11.14 & 0.88 & -9.74 \\
180401.SZ & 2.75 & 0.39 & -3.28 & 4.49 & \textcolor{red}{1.79} & \textcolor{red}{-0.23} & \textcolor{red}{9.37} & 1.08 & -7.98 \\
180501.SZ & \textcolor{red}{28.11} & \textcolor{red}{2.35} & \textcolor{red}{-5.19} & 27.09 & 2.30 & -5.46 & 26.44 & 1.53 & -14.79 \\
180801.SZ & 0.97 & -0.18 & -1.10 & \textcolor{red}{2.06} & \textcolor{red}{0.41} & \textcolor{red}{-0.81} & -4.91 & -0.83 & -7.12 \\
508000.SH & \textcolor{red}{20.40} & \textcolor{red}{2.19} & -5.22 & 18.21 & 2.12 & \textcolor{red}{-4.75} & 11.53 & 0.93 & -9.10 \\
508001.SH & 11.02 & 1.58 & -3.93 & \textcolor{red}{11.12} & \textcolor{red}{1.68} & \textcolor{red}{-3.89} & 7.96 & 0.69 & -10.49 \\
508006.SH & 8.01 & 1.00 & \textcolor{red}{-3.96} & \textcolor{red}{8.04} & \textcolor{red}{1.02} & -3.96 & 7.38 & 0.70 & -5.66 \\
508008.SH & \textcolor{red}{5.73} & 0.74 & -3.47 & 5.54 & \textcolor{red}{0.85} & \textcolor{red}{-2.73} & 3.25 & 0.25 & -9.21 \\
508009.SH & 19.31 & 2.51 & -3.58 & \textcolor{red}{20.22} & \textcolor{red}{2.84} & \textcolor{red}{-2.68} & 14.21 & 1.41 & -9.78 \\
508018.SH & 7.67 & \textcolor{red}{1.06} & -5.15 & 4.70 & 0.81 & \textcolor{red}{-3.95} & \textcolor{red}{10.40} & 1.03 & -8.06 \\
508019.SH & \textcolor{red}{10.85} & 1.46 & -4.66 & 9.40 & \textcolor{red}{1.90} & \textcolor{red}{-3.02} & -13.13 & -1.11 & -24.19 \\
508021.SH & 5.12 & \textcolor{red}{0.56} & -3.93 & 0.44 & -0.13 & \textcolor{red}{-3.43} & \textcolor{red}{6.66} & 0.43 & -12.95 \\
508027.SH & 4.91 & 1.14 & -2.29 & \textcolor{red}{6.94} & \textcolor{red}{1.75} & \textcolor{red}{-1.45} & -8.52 & -0.76 & -17.95 \\
508028.SH & 13.70 & 1.79 & \textcolor{red}{-4.68} & 4.36 & 0.59 & -6.21 & \textcolor{red}{17.66} & \textcolor{red}{1.99} & -5.58 \\
508056.SH & \textcolor{red}{15.19} & 1.98 & -2.41 & 13.32 & \textcolor{red}{2.64} & \textcolor{red}{-1.83} & 7.32 & 0.61 & -7.95 \\
508058.SH & \textcolor{red}{40.23} & \textcolor{red}{3.20} & -7.35 & 35.72 & 3.12 & \textcolor{red}{-6.72} & 38.03 & 2.37 & -16.11 \\
508066.SH & \textcolor{red}{21.58} & 1.96 & -6.02 & 19.41 & \textcolor{red}{2.22} & \textcolor{red}{-5.62} & 13.51 & 0.98 & -11.69 \\
508068.SH & 26.63 & 2.40 & -8.07 & 30.58 & \textcolor{red}{3.07} & \textcolor{red}{-4.81} & \textcolor{red}{31.80} & 2.06 & -12.20 \\
508077.SH & 31.04 & 2.82 & \textcolor{red}{-6.94} & \textcolor{red}{34.37} & \textcolor{red}{3.10} & -8.24 & 30.83 & 1.97 & -14.02 \\
508088.SH & 15.07 & 1.37 & -6.66 & 10.84 & 1.03 & \textcolor{red}{-5.65} & \textcolor{red}{25.68} & \textcolor{red}{1.81} & -8.11 \\
508096.SH & 14.33 & \textcolor{red}{2.44} & \textcolor{red}{-2.83} & 10.42 & 2.02 & -2.84 & \textcolor{red}{17.31} & 1.37 & -9.47 \\
508098.SH & \textcolor{red}{50.25} & \textcolor{red}{4.47} & \textcolor{red}{-3.20} & 37.53 & 3.55 & -4.79 & 49.52 & 3.46 & -5.72 \\
508099.SH & 0.62 & -0.27 & -1.37 & \textcolor{red}{4.16} & \textcolor{red}{0.93} & \textcolor{red}{-1.19} & -8.88 & -0.85 & -13.05 \\
\bottomrule
\end{tabular}

\subsection{Aggregate Net Asset Value Trend Analysis}
To further evaluate the dynamic performance of the strategies from a temporal perspective, Figure 2 presents a comparison of the aggregated total net asset value trends across all fund accounts under the three strategies during the backtest period. The figure shows that the two agent strategies exhibit overall smoother upward trajectories throughout the sample period, with significantly smaller drawdowns during multiple phases of market volatility compared to the benchmark strategy. Specifically, the Agent-DeepSeek-R1 path (blue line) demonstrates a more substantial net value increase during major upward trends and maintains a relatively stable high level in the latter half of the sample. The Agent-Qwen3-8B-FT path (purple line), while slightly lower overall than the blue line, also maintains good smoothness and drawdown control. In contrast, the baseline Buy \& Hold strategy (red dashed line) experiences more pronounced drawdowns in both the initial and later stages, leading to a decline in net value levels. This reflects its greater susceptibility to periodic shocks when no risk exposure adjustments are made. The figure is consistent with the statistical results presented earlier: the multi-agent strategies not only enhance cumulative returns but also significantly improve drawdowns and risk-adjusted returns. However, the figure also reveals limitations in the agent strategies' performance during certain market phases. For instance, during the three rapid market corrections at the end of April, May, and August 2025, the agent strategies also experienced synchronous, substantial drawdowns. Furthermore, they failed to fully capture trend acceleration opportunities during the rising market of March–April 2025. This indicates that the current system's adaptability and trend-capturing capability in highly volatile environments still have room for improvement.

\begin{figure}[H]
\centering
\includegraphics[width=\textwidth]{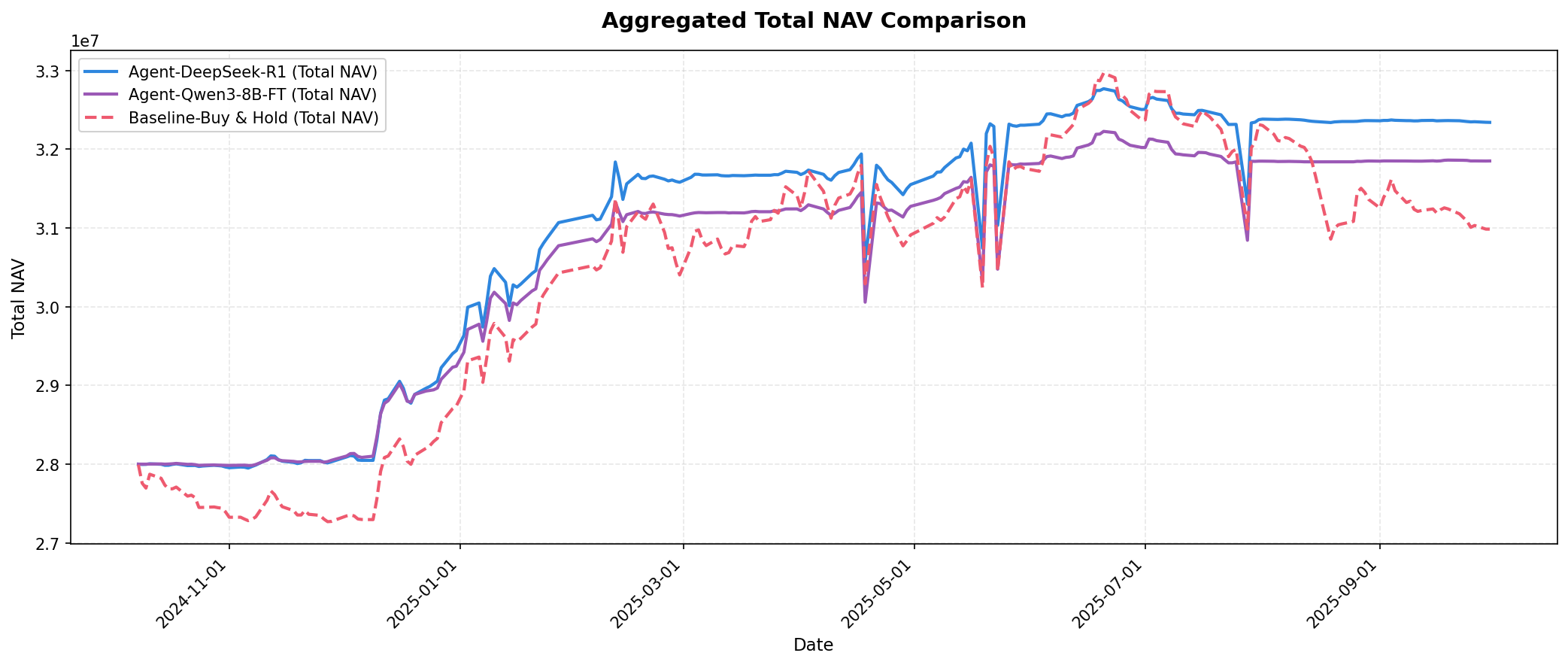}
\caption{Comparison of Aggregate Total NAV across All Fund Accounts for the Three Strategies}
\label{fig:nav_comparison}
\end{figure}
\textbf{Figure Note:} The figure shows the aggregate net asset value curves obtained by summing the net asset values of all single-asset accounts for the sample REITs on a daily basis. The blue line represents the Agent-DeepSeek-R1 pathway, the purple line represents the Agent-Qwen3-8B-FT pathway, and the red dashed line represents the benchmark Buy \& Hold. The horizontal axis is the trading date, and the vertical axis is the aggregate net asset value (Total NAV). This figure is used to depict the overall return paths and drawdown pattern differences of the strategies across the entire market sample.

\section{Future Work}
Although the multi-agent trading framework constructed in this study has demonstrated good return and risk control capabilities in infrastructure public REITs trading, there are still directions for further optimization and expansion:

1. Specialized Training for the Decision Agent

The Decision Agent in the current system employs a general-purpose large language model (DeepSeek-R1) for execution. Its advantage lies in powerful reasoning capabilities and high compliance with prompts. However, this model's decision style is relatively conservative, thus requiring prompt framework constraints. Future work could draw on the training path of the Prediction Agent to construct a specialized fine-tuned model for the decision layer. Through instruction fine-tuning (SFT) and reinforcement learning (e.g., GRPO/DPO, etc.) on historical data, train a small-scale dedicated decision model to enhance decision adaptability and proactiveness.

\nocite{*}
\bibliographystyle{plain}
\bibliography{references}

\end{document}